\documentclass[aps,prl,twocolumn,groupedaddress,showpacs]{revtex4}

\usepackage{graphics}
\usepackage{epsfig}
\usepackage{amsmath}

\begin{document}



\title{Secure Deterministic Communication Without
Entanglement}

\author{Marco Lucamarini$^{1}$}

\author{Stefano Mancini$^{2}$}


\affiliation{$^{1}$Dipartimento di Fisica, Universit\`{a}
di Roma ``La Sapienza", I-00185 Roma, Italy.\\
$^{2}$Dipartimento di Fisica, Universit\`{a} di Camerino, I-62032
Camerino, Italy.}

\date{\today}

\begin{abstract}
We propose a protocol for deterministic communication that does
not make use of entanglement. It exploits nonorthogonal states in
a two-way quantum channel attaining significant improvement of
security and efficiency over already known cryptographic
protocols. The presented scheme, being deterministic, can be
devoted to direct communication as well as to key distribution,
and its experimental realization is feasible with present day
technology.
\end{abstract}


\maketitle

In a recent paper Bostr\"{o}m and Felbinger \cite{bf} presented a
scheme to have ``Secure Deterministic Communication using
Entanglement''. The idea is an original revisiting of the
``Quantum Dense Coding'' \cite{ben1}. They called the protocol
``Ping-Pong" (PP) for its peculiarity of a ``forward and backward"
use of the quantum channel. With respect to Bennett and Brassard's
scheme (BB84) \cite{bb}, the main advantage of PP is its {\it
deterministic} nature, that permits \textit{Quantum Direct
Communication} (QDC) as well as \textit{Quantum Key Distribution}
(QKD). Unfortunately it was proved to be not completely secure
\cite{qi,woj}.

\begin{figure}
\begin{center}
\includegraphics[width=0.45\textwidth]{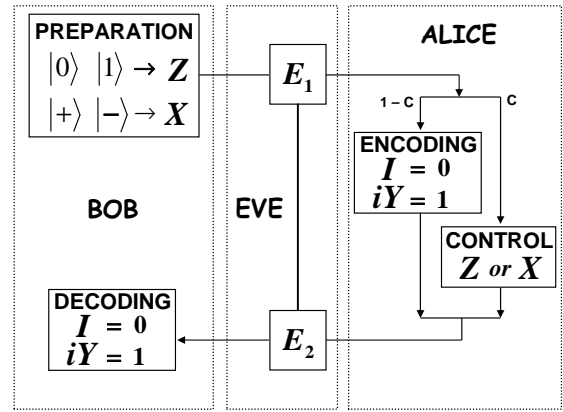}
\end{center}
\caption{\label{Fig1} PP84 scheme}
\end{figure}

We describe here a communication protocol that combines the main
advantages of PP and BB84, while avoiding their drawbacks, and we
termed it PP84 \cite{qi2}. The general idea is summarized in
Fig.\ref{Fig1}. A character, traditionally called Bob, prepares a
qubit in one of the four randomly chosen states $|0\rangle $, $|
1\rangle $ (eigenstates of Pauli operator $Z$), $| +\rangle $, $|
-\rangle $ (eigenstates of Pauli operator $X$), and sends it to
his counterpart Alice. With probability $(c)$ Alice measures the
prepared state (control mode) or, with probability $(1-c)$, she
uses it to encode a bit (encoding mode). After that she will send
the qubit back to Bob. Encoding is represented by a transformation
on qubit state rather than by qubit state itself: identity
operation $I$ encodes $0$, while operation $iY=ZX$ encodes $1$.
Notice that $iY$ acts as a spin-flip on all the beginning states:
\begin{equation}\label{Cod}
\begin{array}
[c]{c} iY\left( \left| 0\right\rangle ,\left| 1\right\rangle
\right) =\left(
-\left| 1\right\rangle ,\left| 0\right\rangle \right) \\
iY\left( \left| +\right\rangle ,\left| -\right\rangle \right)
=\left( \left| -\right\rangle ,-\left| +\right\rangle \right)
\end{array}
\end{equation}
In this way Alice does not need to know the incoming state to
perform the encoding. In turn Bob can easily decode Alice's
message by measuring the qubit in the same basis he prepared it.
This feature makes the protocol deterministic: no qubits are
discarded because of a wrong choice of the basis. To make a QDC it
suffices that Alice performs transformations according to the
message she wants to send out, while for a QKD the sequence of
transformations will be random. Moreover the absence of any public
discussion concerning qubit basis in the
\textit{encoding-decoding} procedure gives an eavesdropper (Eve)
no classical information about qubit state, preventing powerful
attacks based on this strategy.

To guarantee security of PP84, Alice has to measure the qubit with
a probability $c\neq0$ by randomly choosing a basis ($Z$ or $X$),
as in BB84.
After that she will send qubit back to Bob. He, after publicly
declaring the receipt of the qubit, measures it as exactly would
do if Alice had encoded it (actually, he doesn't still know
Alice's choice). At this point Alice reveals publicly whether she
measured (and in which basis) or not, and a public debate on
results is settled with Bob in the former case. Notice that if Eve
is not on the line a perfect ``double" correlation (on the forward
and backward paths) of measurement results must be found by
legitimate users.

The ``run-by-run" protocol just described realizes a QDC, which is
more demanding on security. QKD easily follows from it by
deferring all public discussions at the end of the whole
transmission.

Let us introduce the question of security of PP84 by a simple
eavesdropping strategy. Suppose Eve randomly decides a basis, $Z$
or $X$, along which performing projective measurements on the
traveling qubit, both on forward and backward path. She can guess
the right basis with probability $1/2$, and in this case she is
not revealed at all. If otherwise Eve chooses the wrong basis she
still has a probability of 50\% to evade detection at point
$E_{1}$ and 50\% at point $E_{2}$ giving in the whole a $25\%$ of
possibility to evade detection. This means that Alice and Bob's
double test reveals Eve with average probability equal to: $(0 +
3/4)/2 = 3/8= 37.5\%$. Remarkably, this value is greater than the
one obtained in BB84 (25\%) for the same eavesdropping strategy.
To extend the argument to general attacks we follow the line
sketched in \cite{gis} but with some relevant differences.

Given the four states prepared by Bob, and Eve's ancillary states
$|\varepsilon\rangle$, we can write the most general operation Eve
can do on traveling qubit as:
\begin{align}
\left| 0\right\rangle \left| \varepsilon\right\rangle &
\rightarrow\left| 0\right\rangle \left|
\varepsilon_{00}\right\rangle +\left| 1\right\rangle \left|
\varepsilon_{01}\right\rangle =\sqrt{F}\left| 0\right\rangle
\left| \widetilde{\varepsilon}_{00}\right\rangle +\sqrt{D}\left|
1\right\rangle \left| \widetilde{\varepsilon}_{01}\right\rangle
\nonumber\\
\left| 1\right\rangle \left| \varepsilon\right\rangle &
\rightarrow\left| 0\right\rangle \left|
\varepsilon_{10}\right\rangle +\left| 1\right\rangle \left|
\varepsilon_{11}\right\rangle =\sqrt{D}\left| 0\right\rangle
\left| \widetilde{\varepsilon}_{10}\right\rangle +\sqrt{F}\left|
1\right\rangle \left| \widetilde{\varepsilon}_{11}\right\rangle
\nonumber\\
\left| +\right\rangle \left| \varepsilon\right\rangle &
\rightarrow \frac{1}{\sqrt{2}}\left[ \left| 0\right\rangle \left(
\left| \varepsilon_{00}\right\rangle +\left|
\varepsilon_{10}\right\rangle \right) +\left| 1\right\rangle
\left( \left| \varepsilon_{01}\right\rangle +\left|
\varepsilon_{11}\right\rangle \right) \right] \nonumber\\
& \equiv \left| +\right\rangle \left|
\varepsilon_{++}\right\rangle +\left| -\right\rangle \left|
\varepsilon_{+-}\right\rangle
\nonumber\\
\left| -\right\rangle \left| \varepsilon\right\rangle &
\rightarrow \frac{1}{\sqrt{2}}\left[ \left| 0\right\rangle \left(
\left| \varepsilon_{00}\right\rangle -\left|
\varepsilon_{10}\right\rangle \right) +\left| 1\right\rangle
\left( \left| \varepsilon_{01}\right\rangle -\left|
\varepsilon_{11}\right\rangle \right) \right] \nonumber\\
& \equiv\left| +\right\rangle \left| \varepsilon_{-+}\right\rangle
+\left| -\right\rangle \left| \varepsilon_{--}\right\rangle
\label{Tr1}
\end{align}
Ancillary states are neither orthogonal nor normalized; states
with tilde are instead normalized. The following conditions make
the transformations (\ref{Tr1}) unitary:
\begin{align}
\left\langle \varepsilon_{00}| \varepsilon_{00}\right\rangle
+\left\langle \varepsilon_{01}| \varepsilon_{01}\right\rangle
& \equiv F+D=1 \nonumber \\
\left\langle \varepsilon_{10}| \varepsilon_{10}\right\rangle
+\left\langle \varepsilon_{11}| \varepsilon_{11}\right\rangle
& \equiv D+F=1\nonumber \\
\left\langle \varepsilon_{00}| \varepsilon_{10}\right\rangle
+\left\langle \varepsilon_{01}| \varepsilon_{11}\right\rangle & =0
\label{c3}
\end{align}
We can set, without loss of generality: $\left\langle
\varepsilon_{00}| \varepsilon_{01}\right\rangle =\left\langle
\varepsilon_{10}| \varepsilon_{11} \right\rangle =\left\langle
\varepsilon_{00}| \varepsilon_{10}\right\rangle =\left\langle
\varepsilon_{01}| \varepsilon_{11} \right\rangle =0$. Furthermore,
we specify the angles between nonorthogonal vectors as:
\begin{equation}\label{xy}
   \left\langle \widetilde{\varepsilon
}_{00}|\widetilde{\varepsilon }_{11}\right\rangle =\cos x\,,
\qquad
   \left\langle \widetilde{\varepsilon
}_{01}|\widetilde{\varepsilon }_{10}\right\rangle =\cos y
\end{equation}
with $0\le x\,,\,y\le\pi/2$. Here, we don't fix values of the
parameters we introduced. In \cite{gis} symmetry arguments lead to
assume $F=\left\langle \varepsilon _{00}| \varepsilon
_{00}\right\rangle =\left\langle \varepsilon _{++}| \varepsilon
_{++}\right\rangle$. This reasoning is not applicable in our case:
the absence of a classical discussion during the encoding-decoding
procedure prevents Eve from waiting for the basis revelation, thus
forcing her to break the symmetry by deciding the measurement
basis.

At point $E_{2}$ Eve performs an attack similar to that at point
$E_{1}$, but with fresh ancillae $|\eta\rangle$ (hence new
parameters $F'$ and $D'$),
\begin{align}\label{Tr2}
   \left| 0\right\rangle \left| \eta \right\rangle
&\rightarrow
   \sqrt{F^{\prime
}}\left| 0\right\rangle \left| \widetilde{\eta }_{00}\right\rangle
+\sqrt{D^{\prime }}\left| 1\right\rangle \left| \widetilde{\eta
}_{01}\right\rangle \nonumber\\
   \left| 1\right\rangle \left| \eta \right\rangle
&\rightarrow
   \sqrt{D^{\prime }}\left| 0\right\rangle
\left| \widetilde{\eta }_{10}\right\rangle +\sqrt{F^{\prime
}}\left| 1\right\rangle \left| \widetilde{\eta
}_{11}\right\rangle \nonumber \\
   \left| +\right\rangle \left| \eta
\right\rangle &\rightarrow \left| +\right\rangle \left| \eta
_{++}\right\rangle +\left| -\right\rangle \left| \eta
_{+-}\right\rangle \nonumber \\
   \left| -\right\rangle
\left| \eta \right\rangle &\rightarrow \left| +\right\rangle
\left| \eta _{-+}\right\rangle +\left| -\right\rangle \left| \eta
_{--}\right\rangle
\end{align}
At the end of transmission Eve will measure $\varepsilon$ and
$\eta$ ancillae and, by comparing results, she will gain
information. This strategy represents the most general attack Eve
can perform in a single run, if we do not allow her to create
coherence in some way between forward and backward paths; for this
reason we term it ``incoherent", referring to the other
possibility as ``coherent". Within this kind of attack we must
recover the optimal measure Eve can do i.e. we must maximize Alice
and Eve's mutual information (${\cal I}_{AE}$) minimizing
detection probability ($P_{d}$).

From transformations (\ref{Tr1}) and conditions (\ref{c3}) we can
evaluate the probability that Eve {\it is not detected} in the
forward path, after her $E_{1}$-attack:
\begin{align}\label{notd}
P_{nd}(\left| 0\right\rangle ) & =\left\langle
\varepsilon_{00}| \varepsilon_{00}\right\rangle =F\\
P_{nd}(\left| 1\right\rangle ) & =\left\langle \varepsilon_{11}|
\varepsilon_{11}\right\rangle
=F\nonumber\\
P_{nd}(\left| +\right\rangle ) & =\left\langle \varepsilon_{++}|
\varepsilon_{++}\right\rangle =(1/2)\left[
1+F\cos x+D\cos y\right] \nonumber \\
P_{nd}(\left| -\right\rangle ) & =\left\langle \varepsilon_{--}|
\varepsilon_{--}\right\rangle=(1/2)\left[ 1+F\cos x+D\cos y\right]
\nonumber
\end{align}
Similar arguments hold for the backward path, after
$E_{2}$-attack, with primed parameters replacing not-primed ones.
The probability that Eve is not detected after a whole run is then
the product of the two partial probabilities; by taking its
complement we obtain the probability to detect Eve. Averaging it
over all input states we get:
\begin{align}\label{Pav}
\nonumber P_{d}& = (1/8)\{ 7-4FF^{\prime }-F\cos x-D\cos
y-F^{\prime }\cos x^{\prime }
\\
&-D^{\prime }\cos y^{\prime} -FF^{\prime }\cos x\allowbreak \cos
x^{\prime} -FD^{\prime }\cos x\cos y^{\prime }
\nonumber \\
&-D\allowbreak F^{\prime }\cos y\cos x^{\prime }-DD^{\prime }\cos
y\cos y^{\prime }\}
\end{align}
It is possible to show \cite{dc} that $P_{d}$ takes the minimum
\begin{equation}\label{Pmin}
d\equiv\min P_{d} = \left[ 1-\left( 1+\cos x\right) \left( 1+\cos
x^{\prime }\right)/4 \right]/2
\end{equation}
for $F=F'=1$; this condition represents the best Eve can do to
conceal her presence. The maximum value $d=3/8$ is obtained when
$x=x'=\pi/2$, corresponding, as we will see, to Eve's maximum
information.

If Alice and Bob are going to make a QDC we can evaluate the
probability that Eve can steal $n$ bits of full information
without being detected as $(1-c)^n/[1-c( 1-d)]^{n}$. It turns out,
for example, that if $c=1/2$ Eve has a probability of about
$7.8\%$ to successfully eavesdrop 1 byte (i.e. 8 bits) of full
information and about $0.6\%$ to eavesdrop 2 bytes. Increasing the
value of $c$ increases security of the protocol, but at the
expense of the transmission rate.

If instead Alice and Bob are going to make a QKD then the argument
is a little bit more complicated. To evaluate ${\cal I}_{AE}$ let
us write Bob's initial state as $\left| \Psi \right\rangle
=\sum_{\alpha =0,1}C_{\alpha }\left| \alpha \right\rangle$. Now,
suppose Alice performs identity between the two Eve's attacks:
\begin{eqnarray}\label{I}
&& \left| \Psi \right\rangle \left| \varepsilon \right\rangle
\left| \eta \right\rangle \overset{E_{1}}{\rightarrow }
     \sum_{\alpha }C_{\alpha }\sum_{\beta } \left|
     \beta \right\rangle \left| \varepsilon _{\alpha
\beta }\right\rangle \left| \eta
\right\rangle\overset{I}{\rightarrow }
\\
&&\sum_{\alpha }C_{\alpha }
     \sum_{\beta } \left|
\beta \right\rangle \left| \varepsilon _{\alpha \beta
}\right\rangle \left|
     \eta \right\rangle \overset{E_{2}}{\rightarrow }
\sum_{\alpha }C_{\alpha }\sum_{\beta ,\gamma } \left| \gamma
\right\rangle \left| \varepsilon _{\alpha \beta }
     \right\rangle \left| \eta _{\beta \gamma
}\right\rangle \nonumber
\end{eqnarray}
The ancillary states involved in this operation are:
\begin{align} \label{ISt}
\nonumber
     \left| \varepsilon _{00},\eta _{00}\right\rangle
,\left| \varepsilon _{00},\eta _{01}\right\rangle ,
     \left| \varepsilon _{01},\eta _{10}\right\rangle
,\left| \varepsilon
_{01},\eta _{11}\right\rangle \\
     \left| \varepsilon _{10},\eta _{00}\right\rangle
,\left| \varepsilon _{10},\eta _{01}\right\rangle ,
     \left| \varepsilon _{11},\eta _{10}\right\rangle
,\left| \varepsilon _{11},\eta _{11}\right\rangle
\end{align}
If, instead, Alice performs a flip we have:
\begin{eqnarray}\label{Y}
&& \left| \Psi \right\rangle \left| \varepsilon \right\rangle
\left| \eta \right\rangle \overset{E_{1}}{\rightarrow }
     \sum_{\alpha }C_{\alpha }\sum_{\beta }
     \left| \beta \right\rangle
  \left| \varepsilon _{\alpha
\beta }\right\rangle \left| \eta \right\rangle
\overset{iY}{\rightarrow }\nonumber
\\
&&\sum_{\alpha }C_{\alpha }\sum_{\beta }
     \left( -1\right) ^{\beta +1} \left| \beta
\oplus 1\right\rangle \left| \varepsilon _{\alpha \beta
}\right\rangle\left| \eta \right\rangle
     \overset{E_{2}}{\rightarrow }
\nonumber\\
&& \sum_{\alpha }C_{\alpha }\sum_{\beta ,\gamma }\left( -1\right)
^{\beta +1}
     \left| \gamma \right\rangle\left| \varepsilon
_{\alpha \beta }\right\rangle \left| \eta _{\left( \beta \oplus
1\right) \gamma }\right\rangle
\end{eqnarray}
and ancillary states involved are:
\begin{align} \label{YSt}
\nonumber
     \left| \varepsilon _{00},\eta _{10}\right\rangle
,\left| \varepsilon _{00},\eta _{11}\right\rangle ,
     \left| \varepsilon _{01},\eta _{00}\right\rangle
,\left| \varepsilon
_{01},\eta _{01}\right\rangle \\
     \left| \varepsilon _{10},\eta _{10}\right\rangle
,\left| \varepsilon _{10},\eta _{11}\right\rangle ,
     \left| \varepsilon _{11},\eta _{00}\right\rangle
,\left| \varepsilon _{11},\eta _{01}\right\rangle
\end{align}
To acquire information from states (\ref{ISt}) and (\ref{YSt}) Eve
must measure both her ancillae. Keeping in mind orthogonality
relations (\ref{c3}) and following, we see that the best way to do
that is to distinguish orthogonal subspaces before, and then
nonorthogonal states within them. The probability to correctly
distinguish between two states with scalar product $\cos x$ is
$\left(1+\sin x \right)/2 $ \cite{per}. Observing states
(\ref{ISt}) and (\ref{YSt}) we can notice that if Eve mistakes to
identify her first ancilla ($\varepsilon $ states) then she
guesses wrong Alice's operation, since she flips from states
(\ref{ISt}) to (\ref{YSt}) or viceversa. The same is true if she
guesses right $\varepsilon $ state but mistakes $\eta $ state.
Nevertheless, if she mistakes twice, then with the first error she
misinterprets (\ref{ISt}) with (\ref{YSt}) and with the second
error she compensates the first, eventually guessing right Alice's
operation. This lead to a lengthy expression of ${\cal I}_{AE}$ as
a function of the six parameters describing ancillae states, but
it can be simplified recalling that Eve wants to keep the $P_{d}$
as low as possible, and so condition $F=F'=1$ applies. In this
case Eve's strategy is optimal and ${\cal I}_{AE}$ becomes:
\begin{eqnarray}\label{IAE}
   {\cal I}_{AE}
   = 1-h\left[\left( 1+\sin x\sin x^{\prime
}\right)/2 \right]
\end{eqnarray}
where $h(\cdot)$ indicates the Shannon binary entropy \cite{niel}.

We are now in the position to compare relevant quantities we have
calculated, in particular Eq.(\ref{Pmin}) for $P_{d}$ and
Eq.(\ref{IAE}) for ${\cal I}_{AE}$. Unfortunately, both equations
are functions of $x$ and $x^{\prime}$ thus preventing us to write
${\cal I}_{AE}$ as a function of $P_{d}$. However, the following
lemma holds:

\textbf{Lemma}.\emph{The optimal Eve incoherent attack consists in
a {\it balanced} one for which $x=x^{\prime}$.}

This lemma can be justified with the following qualitative
argument \cite{dc}. The orthogonality Eve imposes on her ancillae
is somewhat related to the information she can extract from the
qubit: the more orthogonal they are, the higher is the information
gained. If she sets $x>x^{\prime}$, the ancillae $\varepsilon$
will be more orthogonal than ancillae $\eta$, and this entails a
loss of information when going from the forward to the backward
path. If she sets $x<x^{\prime}$ we can argue the reverse.

\begin{figure}
\begin{center}
\includegraphics[width=0.45\textwidth]{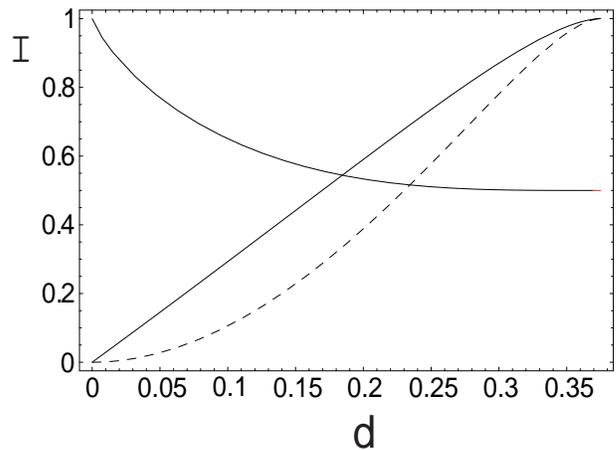}
\end{center}
\caption{\label{Fig4} PP84 Security: Information vs Detection
Probability. The descent curve represents Alice and Bob mutual
information ${\cal I}_{AB}$. The crescent curves represent Alice
and Eve mutual informations (${\cal I}_{AE}$ dashed line, ${\cal
I}_{AE}^{bound}$ solid line).}
\end{figure}

The above lemma allows us to write $P_{d}$ (\ref{Pmin}) as:
\begin{equation}
\label{Pmx} d = (1/2)-(1/8)\left( 1+\cos x\right)^{2}
\end{equation}
By inverting this relation and substituting it into Eq.(\ref{IAE})
we can express the information ${\cal I}_{AE}$ as function of
$P_{d}$:
\begin{equation}\label{IAEd}
{\cal I}_{AE}=1-h\left\{\left[ 2-( 2\sqrt {1-2d}-1)^{2}
\right]\Big/2\right\}
\end{equation}
It is easy to see that the maximum of information, ${\cal
I}_{AE}=1$, corresponds to a detection probability $d=37.5\%$.
This implies that projective attacks described above are a kind of
optimal incoherent attacks. ${\cal I}_{AE}$ as a function of
$P_{d}$ is shown in Fig.\ref{Fig4}.
For a QKD to be secure a comparison between Alice and Eve's mutual
information and Alice and Bob's one must be settled; we must then
evaluate the amount of information Bob has access to. Also in
these calculations we must set $F=F'=1$, because Bob receives a
perturbed state according to Eve's choice of minimizing the
$P_{d}$. So, with this condition, the probability that Bob makes a
right guess on Alice's transformation after preparing $\left|
+\right\rangle $ or $\left| -\right\rangle $, and Eve measures in
the basis $\left| \varepsilon_{0,1}\right\rangle, \left|
\eta_{0,1}\right\rangle$, is $\label{Px} P_{0,1} = \left(
\allowbreak1+\cos x\cos x^{\prime} \right)/2 $. The probability
that Bob makes a right guess on Alice's transformation, after
preparing the same states and Eve measures in the basis $\left|
\varepsilon_{+,-}\right\rangle, \left| \eta_{+,-}\right\rangle$ is
$\label{Pz} P_{+,-}=1 $. Analogous results hold if Bob prepares
states $\left| 0\right\rangle $ or $\left| 1\right\rangle $.
Averaging the information corresponding to $P_{0,1}$ and to
$P_{+,-}$, and using the above \textit{Lemma} we get:
\begin{equation}
{\cal I}_{AB}=1-(1/2)h\left[\left(1+\cos^{2} x\right)/2 \right]
\label{Ibob2}
\end{equation}
Eq.(\ref{Ibob2}) can be written as function of $P_{d}$ by
inverting, as for ${\cal I}_{AE}$, Eq.(\ref{Pmx}). The result is
shown in Fig.\ref{Fig4}. We notice that minimum Bob's information
is $1/2$, because in half cases Eve doesn't perturb the channel at
all. Nevertheless, if we had used symmetry conditions discussed
after Eq.(\ref{xy}) we would have obtained ${\cal I}_{AE}$ going
to zero. The intersection of ${\cal I}_{AB}$ and ${\cal I}_{AE}$
determines the safety of the protocol for QKD. It results that the
PP84 is secure against a general incoherent attack provided
$d\lesssim 23\%$.

So far we have considered incoherent attacks and evaluated
relevant quantities like detection probability, Eve's information
and Bob's information. What is about coherent attacks? The most
general coherent attack can be written as a transformation on 64
states of the form:
\begin{align}
\left| 0\right\rangle \left| \varepsilon_{00}\right\rangle \left|
\eta\right\rangle & \rightarrow\left| 0\right\rangle \left[ \left|
\varepsilon_{00}\right\rangle \left| \eta_{000000}\right\rangle
+\left| \varepsilon_{01}\right\rangle \left|
\eta_{001000}\right\rangle
\right.\nonumber\\
&\left.+\left| \varepsilon_{10}\right\rangle \left|
\eta_{010000}\right\rangle +\left| \varepsilon_{11}\right\rangle
\left| \eta_{011000}\right\rangle \right]
+\nonumber\\
& +\left| 1\right\rangle \left[ \left|
\varepsilon_{00}\right\rangle \left| \eta_{100000}\right\rangle
+\left| \varepsilon_{01}\right\rangle \left|
\eta_{101000}\right\rangle
\right.\nonumber\\
&\left.+\left| \varepsilon_{10}\right\rangle \left|
\eta_{110000}\right\rangle +\left| \varepsilon_{11}\right\rangle
\left| \eta_{111000}\right\rangle \right] \nonumber\\
|1\rangle|\varepsilon_{00}\rangle|\eta\rangle &
\rightarrow\ldots\nonumber\\
&\vdots \label{Coh1}
\end{align}
because the attack at point $E_{2}$ includes also the resulting
state of the first ancillae. This can create coherence between
forward path and backward one, and Eve can take advantage of it.
Despite of that the most general expression for detection
probability remains the one found for incoherent attacks, given by
Eq.(\ref{Pmx}), because it is obtained when Alice decides to
measure, and her measure breaks any coherence Eve could have
created. Moreover, we notice that the first (``translucid"
\cite{ek}) attack at point $E_{1}$ is the most general Eve can
perform on the forward path, and so it must be the same for any
kind of complete attack. This arguments permit us to bound the
information Eve can acquire in any attack, coherent or not. We
simply put together a $P_{d}$ equal to $d$ and a value for ${\cal
I}_{AE}$ equal to the one Eve would obtain by setting $x'=\pi/2$
in Eq.(\ref{IAE}): ${\cal I}_{AE}^{bound}
   = 1-h[(1+\sin x)/2]$. It is clear that we have overestimated Eve's potential information gain or,
the same, underestimated $P_{d}$. Also this result is shown in
Fig.\ref{Fig4}. We obtain in this way the very lower bound for
security of PP84, namely $d\lesssim 18\%$. Notice that it is still
greater than the one found in similar circumstances (individual
attacks on a lossless channel) for BB84 ($d\lesssim 15\%$)
\cite{gis}. We believe that these results, together with the
introductory remarks on attacks based on classical information
exchange during the encoding-decoding procedure, demonstrate the
safety of PP84.

We would also like to briefly describe the behavior of PP84 when
losses are present on the quantum channel. Two aspects of this
question must be addressed: security against losses-based attacks
and efficiency of transmission. As far as the former is concerned
the risk is that an almighty Eve could substitute an imperfect
channel with a perfect one and conceal her attacks behind losses
interpreted as natural by Alice and Bob. This possibility exploits
the lack of symmetry of Alice control (\cite{woj}, \cite{woj2}),
and is not effective in our case. We have also considered more
subtle attacks based on a kind of ``quantum nondemolition
strategy" by Eve, but Alice and Bob can always detect them by
comparing losses rate pertaining to control mode runs and encoding
mode ones.

As far as efficiency of transmission is concerned the implicit
assumption is that losses are not due to an eavesdropper, but only
to an imperfect channel. Accordingly to Ref.\cite{cab} the
theoretical efficiency of a cryptographic protocol for QKD is
given by the formula $\cal{E}$ = $b_{s} / \left(
q_{t}+b_{t}\right) $, where $b_{s}$ is the expected number of
secret bits received by Bob, $q_{t}$ is the number of transmitted
qubits on the quantum channel, and $b_{t}$ is the number of
transmitted bits on the classical channel. Since in PP84 no
classical information is needed in the encoding-decoding procedure
we have $b_{t}=0$, that entails ${\cal E}=1$ (while for BB84
${\cal E}=1/4$). As far as practical efficiency is concerned we
refer to \cite{deg}. In our protocol a qubit, represented for
instance by the polarization of a photon, is traveling for a
distance $2L$, being $L$ the separation between Alice and Bob. If
${\cal P}$ is the probability that a photon is transmitted over
the distance $L$, we have a total probability ${\cal P}^2$ to
transmit it over $2L$. The practical efficiency can then be
evaluated as ${\cal E}'={\cal E}{\cal P}^{2}$. For BB84 the photon
only travel for a distance $L$, thus ${\cal E}'={\cal P}/4$.
Comparing the two protocols, we obtain that if ${\cal P}\geq 25\%$
then the most efficient scheme is just the PP84.

In conclusion, we have presented a cryptographic protocol which
embeds peculiarities of BB84 and PP protocols, allowing direct
communication as well as key distribution. It has been proved
asymptotically secure in the former case and secure against
general attacks in the latter case, admitting a greater quantum
bit error rate with respect to other known protocols. It is more
efficient than other schemes when losses are taken into account,
whilst allowing easier experimental realization.

\textit{Acknowledgements.} We acknowledge useful discussions with
G. Di Giuseppe and S. Pirandola. One of us (M.L.) would like to
thank his parents for funding.




\begin{references}
\bibitem{dc}
The detailed calculations will be presented elsewhere for space
reasons.
\end{references}
\end{document}